# *Relation between the Usual Order* and the *Enumeration Orders* of *r.e. sets* (2) :

# Enumeration Reducibility

**Ali-Akbar Safilian**
(Amirkabir University of Technology (Tehran Polytechnic), Tehran, Iran
ali_safilian@aut.ac.ir)

**Farzad Didehvar**
(Amirkabir University of Technology (Tehran Polytechnic), Tehran, Iran
didehvar@aut.ac.ir)

**Abstract:**
In this article we define a new reducibility based on enumeration orders.

**Keywords:** Listing, Uniformity, Computability Enumeration Order Reducibility.

## 1    Introduction

We are interested in enumeration orders of r.e. sets. One of the first results of this effort was represented on April 2009 on arXiv. It is the article "Relation between the Usual Order and the Enumeration Orders of Elements of r.e. sets".[1] The current article is an effort to change a little bit the first one. In the new point of view, we start from ordering relation as a first step and in the second step we defined the equivalence relation.

In our first attempt in a wrong way we think that the definition of ordering could not be useful. Here, our special thanks for Professor Klaus Weihrauch which mentioned us our fault and also some other valuable comments.

In the first version we have defined "Uniformity" concept on sets and also computable functions and reached some properties of that. In section 2, we have introduced some of them.

In section 3, we introduce only the definitions and some theorems. The other results of this study will be published subsequently.

## 2    Uniformity

In this section we introduce some definitions and results reached in the first version of the study.

**Definition 1 (listing)** A *listing* of an infinite r.e. set $A \subseteq N$ is a bijective computable function $f: N \to A$.

**Definition 2 (Uniformity on listings and sets)**
1. Two listings $h, g$ are uniform, $h \sim g$, if $h(i) < h(j) \Leftrightarrow g(i) < g(j)$ for all $i, j \in N$.
2. Two r.e. subsets $A, B$ of $N$ are *uniform*, $A \sim B$, if there exist listings $h$ of $A$ and $g$ of $B$ such that $h \sim g$.

**Theorem 3** For r.e. sets $A, B$ the following statements are equivalent:
1. $A \sim B$,
2. For every listing $h$ of $A$ there is a listing $g$ of $B$ such that $h \sim g$.

**Proposition 4** If two sets belong to same uniformity equivalence class, then they do not belong necessarily to same 1-1 reducibility equivalence class.□

**Proposition 5** If two r.e. sets $A$ and $B$ are uniform then they belong to the same Turing-reducibility equivalence class.

**Theorem 6** There is infinite number of uniformity equivalence classes which are located in any Turing degree equivalence class.

In addition to uniformity, we defined another concept named "Type-2 Uniformity" and studied it.
Since almost everything for finite sets in trivial, we only consider infinite r.e. sets.

## 3 Enumeration Reducibility

In this section, we define a reduction named "E-Reducibility "on both natural number function and sets. And then try to get the concept uniformity by means of it.

**Definition 7 (E-Reducibility on listings)**
For listings $f, g: N \to N$ we say $f <_{eo} g$ if and only if $\forall i < j \ f(i) < f(j) \Rightarrow g(i) < g(j))$

**Definition 8 (E-Reducibility on infinite subsets of natural numbers)**
For r.e. sets $A, B \subseteq N$, we say $A <_{eo} B$ if and only if there exist two listings $f, g$ for $A$ and $B$ respectively such that $f <_{eo} g$

**Definition 9 (Uniform Enumeration Orders on listings and sets)**
1. For two computable functions $f, g: N \to N$ we say $f \approx_{eo} g$ if and only if $f <_{eo} g$ and $g <_{eo} f$,
2. For two r.e. sets $A, B \subseteq N$, we say $A \approx_{eo} B$ if and only if $A <_{eo} B$ and $B <_{eo} A$.

Now, our aim is to develop these concepts to reach some valuable results. In the following, we introduce some results on enumeration reducibility.

**Theorem 10.** $\approx_{eo}$ and $\sim$ are equivalnet.

**Proof**: ….

**Conclusion 11.** There exist infinite number of sets $A_1$, $A_2$, ... such that $[K]_{eo} < [A_1]_{eo} < [A_2]_{eo} < \cdots < [\emptyset]_{eo}$

**Conclusion 12** There exist infinite number of sets $B_1, B_2, ...$ such that are not comparable by E-Reducibility.